# A text-independent speaker verification model: A comparative analysis

Rishi Charan, Manisha.A, Karthik.R, Rajesh Kumar M, *Senior IEEE Member*
School of Electronic Engineering
VIT University
Tamil Nadu, India
rishicharan96@gmail.com , manishaab26@gmail.com, mrajeshkumar@vit.ac.in

*Abstract*—The most pressing challenge in the field of voice biometrics is selecting the most efficient technique of speaker recognition. Every individual's voice is peculiar, factors like physical differences in vocal organs, accent and pronunciation contributes to the problem's complexity. In this paper, we explore the various methods available in each block in the process of speaker recognition with the objective to identify best of techniques that could be used to get precise results. We study the results on text independent corpora. We use MFCC (Mel-frequency cepstral coefficient), LPCC (linear predictive cepstral coefficient) and PLP (perceptual linear prediction) algorithms for feature extraction, PCA (Principal Component Analysis) and t-SNE for dimensionality reduction and SVM (Support Vector Machine), feed forward, nearest neighbor and decision tree algorithms for classification block in speaker recognition system and comparatively analyze each block to determine the best technique.

*Index Terms*— MFCC (Mel-frequency cepstral coefficient), LPCC (linear predictive cepstral coefficient), PCA (Principal component analysis) and t-SNE.

## I. Introduction

Speaker verification is considered one of the essential biometric methods in assuring identity in numerous real world applications [1]. Speaker recognition is actually identifying an individual's voice from a set of potential speakers while verification is confirming a speaker's identity as the original speaker or as a trespasser who could be trying to intrude. In this paper speaker identification is the area of interest. The Speaker identification technique has three main operations which are: Feature Extraction, dimensionality reduction and classification. *Feature Extraction:* Voice signal is converted into a set of 12-15 features or feature vectors for further proceedings in the model. *Dimensionality reduction*: This module is used to lower the dimensions of the extracted feature set which makes implementation of the classification techniques easier. *Classification:* This module is useful in multi-speaker recognition problems. The given voice signal is segmented into equal length voice segments and labels are assigned to identify the speaker.
Md Raibul et al [2] have already worked on speaker identification which uses cepstral features and PCA for classification. An enhance study was done by Muda.L [3] et al on MFCC and dynamic time warping techniques to obtain a better performance. Urmila Shrawankar et al and MJ Alam et al [4, 5, 6] have also done an extensive analysis on feature extraction methods like MFCC, PLP, FFT, LPC and LPCC etc.

Our paper aims to bring out a comparative analysis on each module and also to determine the most efficient combination of algorithms that could be used to obtain a reliable outcome. Feature extraction is one of the most widely researched areas when it comes to speaker recognition. State of art methods like MFCC and hidden Markov model have been studied extensively for more than a decade now. But in this paper, we have implemented three different for feature extraction techniques namely MFCC, LPCC and PLP. In dimensionality reduction module, our work focuses on two popular techniques: PCA and t-SNE (Stochastic neighbour embedding). The last module compares different classifiers such as nearest neighbour, SVM, Feed forward network and decision tree. Results of each module are compared individually as well as sequentially to decipher the best way to recognize a speaker.

## II. RELATED WORK

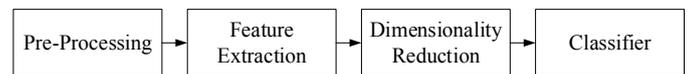

Fig.1: Block diagram of speaker verification model

*A. Pre-Processing:*

In pre-processing, we are going to remove maximum part of silence present in the signal, for achieving this we are going to use theory of probability density function to remove noise and silence part of the signal. Usually first 200ms of any recorded voice signal corresponds to the silence as there is always a time gap between the point where the speaker starts talking and the voice starts to be record this time is habitually minimum of 200ms. Normal density function is used to remove silence and to find the endpoints of the signal. A one-dimensional Gaussian distribution has 68% of its probability mass in the range |u|≤1, 95% in the range of |u|≤2, and 99.7% in the range of |u|≤3. Where u is defined as follows

$$u = \frac{x-\mu}{\sigma} \quad (1)$$

Where μ, σ are the mean and variance of the first 200ms of the speech signal. Algorithm figure 1.3 was used to discriminate between voice part of signal from unvoiced part of the signal. Pre-processing of Speech Signal serves various purposes in any speech processing application. It includes Noise Removal, Endpoint Detection etc. figure 1.1, figure 1.2 shows the Input and output of the Pre-Processing blocks.

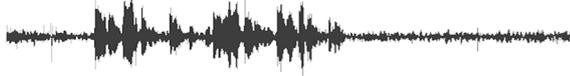

Figure1.1: Input from the microphone

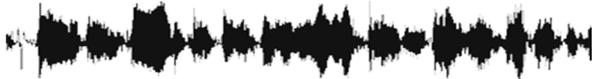

Figure 1.2: Output signal of Pre-processing

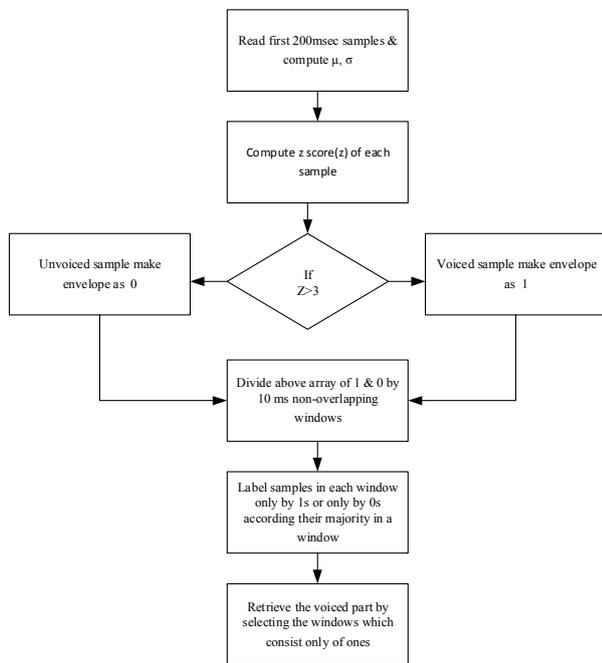

Figure 1.3: Algorithm used for end point detection and silence removal (pre-processing)

### B. Feature Extraction:

The voice algorithms consist of two parallel paths. The first one is training sessions, in this part we feed the voice signals along with their identity to the algorithm so that the extracted features can be categorized the second one is categorized as testing where this is the one which is used for identification of the individual. In voice identification feature extraction plays an important role in extracting the features from the infinite information containing voice signal which can be used for identifying the speaker among a group of N number of speakers. We are going to use MFCC, LPCC and PLP techniques for extraction of Short-term spectral features which will be compared to find the best possible extraction method for different applications. Voice signals are non-stationery for a large duration and stationery when we take them for a short duration of 20-25msec duration. We use these techniques for extraction of these stationery features.[8]

*1. Mel Frequency Cepstral Coefficients (MFCC):*

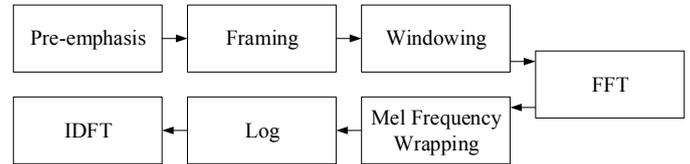

Figure 2.1: complete pipeline of MFCC

MFCC uses all-zero model for computing spectra. The output of pre-processing block is taken as input to the feature extraction stage where pre-emphasis is done to the signal to increase the energy of the signal at higher frequencies as it also removes DC offset present in the signal. Transfer function of this step is as follows

$$Y[n] = x[n] - a*x[n-1] \quad (2)$$

Where value of a lies in between [0.9,1], from the figure 2.2.1 and figure 2.2.2 we can see the central frequency of the speaker along with signal strength at higher frequency that is changing when 'a' value is changed from 0.9 to 1

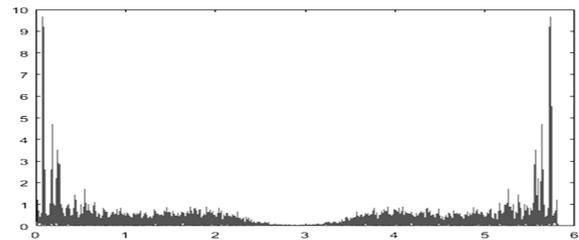

Figure 2.1.1 FFT of the signal when a=1

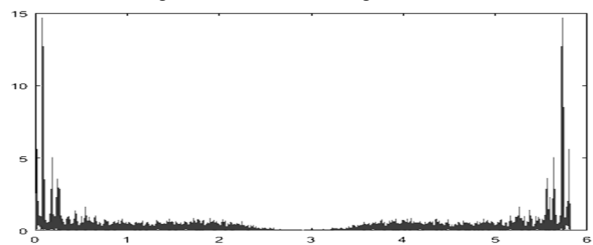

Figure 2.1.2 FFT of the signal when a=0.9

Fig 2.1.1 and Fig 2.1.2 shows the FFT of the given audio signal where x-axis represents the frequency and y-axis represents the amplitude when 'a' is equal to 1 and 0.9 respectively. Now the signal is divided into a set of short frames with a duration of 20-25ms as voice signal which is considered to have stationery features for short period of time with each frame having an overlap region of 50-80% with the other frames, we use windows most preferably a window to decrease the strength of the samples at the end of the frame. Commonly used windows are Hamming, Hanning, Blackman, Rectangular and Triangular windows.

Hamming: $-w(n) = 0.54 - 0.46 * \cos\frac{2\pi n}{N-1}$ (3)

We consider hamming window for windowing process. FFT is found for individual frames and they are passed through Mel frequency bank.

Mel(f)=$1125*\ln\left(1 + \frac{f}{700}\right)$ (4)

In Mel frequency wrapping we multiply FFT of the frames with their Mel bank values and output logarithm is sent to IDCT (Inverse discrete cosine transformation) to get the desired number of features. Then logarithm is applied, as it compresses dynamic range of values as human responses are logarithmic to signal responses. Figure 2.1.3 shows features for all the frames. [3, 4, 5]

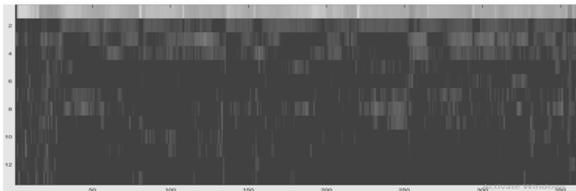

Figure 2.1.3 MFCC features of the signal where x-axis is frame number

*2.Linear Prediction Cepstral coefficients:*

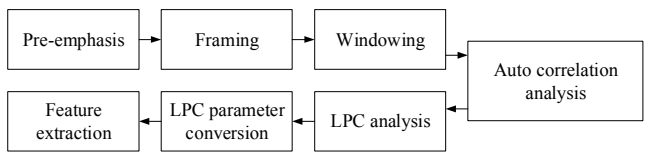

Figure:2.2.1: complete pipeline of LPCC

In LPCC we are going to use all-pole or maximum entropy model or auto regression model for calculation of the spectra which is counter part of all-zero model used in MFCC extraction. first Linear Predictive Coding (LPC)[9] coefficients are found and then they are converted to cepstral coefficients . LPCC is also a well-known algorithm and widely used to extract feature in speaker signal. LPC parameters effectively describe energy and frequency spectrum of voiced frames. The base of explaining acoustic signals spectrum, modeling and pattern recognition is set by the result of increasing logarithm which restrains the fast change of frequency spectrum, more centralized and better for short-time character and it is because of Cepstrum derived from original spectrum. One of the common short-term spectral measurements currently used are LPC derived cepstral coefficients (LPCC) and their regression coefficients. Order Q of auto regression model used for computation of LPC is the number of concentric cylinders used to model the vocal track where $8 \leq Q \leq 16$ figure 2.2.1 shows algorithm of LPCC feature extraction LPCC extracted features of the audio is shown in the figure 2.2.2 with frames along x-axis. [4, 5, 6, 7]

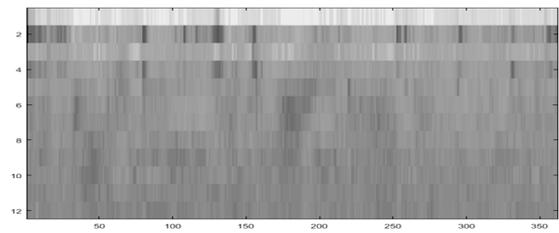

Figure 2.2.2 LPCC features of audio signal x-axis is frames

*3.Perceptual linear prediction*

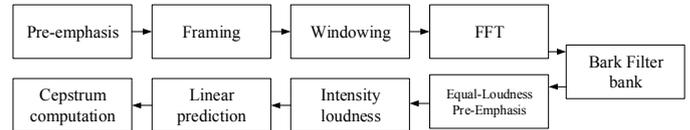

Figure 2.3.1: complete pipeline of PLP

PLP is an extended version of linear prediction coefficients till Fast Fourier transformation (FFT) , same procedure is followed as described in MFCC feature extraction. PLP is combination of concepts of LPCC and MFCC for computation of coefficients PLP uses Bark scale instead of Mel scale like in MFCC feature extraction.

Bark=$6 * \ln\left(\frac{\omega}{1200\pi} + \left(\left(\frac{\omega}{1200\pi}\right)^2 + 1\right)^{\frac{1}{2}}\right)$ (5)

Next step Equal loudness pre emphasis is designed to do some pre-emphasis in the spirit of combining the concept of Equal Loudness Curves.it is a process to normalize different loudness in the voice frames. We calculated Intensity-loudness power is found from the output of equal power pre emphasis by taking cubic root of the Equal loudness pre emphasis[4, 5,7]. Till this we have taken concepts of MFCC, from here we use next half of LPCC like finding LPC coefficients followed by spectrum converted to Cepstrum figure 2.3.2 shows the extracted PLP features of the audio signal[10].

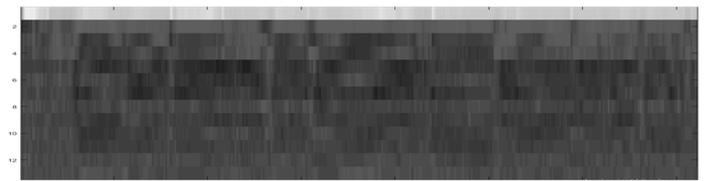

Figure 2.3.2 PLP features extracted for audio signal x-axis is frames

*C. Dimensionality reduction:*

In theoretical point of view more the number of features better is the performance but as number of features increase the performance of the system decreases.so in order to increase the performance of the algorithm we use Dimensionality reduction techniques. Dimensionality reduction means information loss so our main objective in choosing a Dimensionality reduction technique is to preserve as much information as possible while reducing the dimension of the voice signal. In this paper, we are going to use two dimensionality reduction techniques

*1.Principle Component Analysis (PCA):*
A data matrix of n features and m dimensions, which can be correlated can be converted into a matrix of q features, which are non-correlated axis's. Objective of PCA is to rigidly rotate the axes of this m-dimensional space to new position (principal axis) so that n principle axis at in descending order and the co-variances are zero. Principle axis can be found by using eigen analysis of the cross product matrix(s).

$$|S - \lambda I| = 0 \quad (6)$$

Matrix λ is the variance of the coordinates on each principal component axis. Eigen vector with highest Eigen value is the principle component it is the most significant relation between our variable and the dimension.

*Final data=Row Feature Vector X Row Data Adjust* (7)

Row Feature vector is the matrix with the eigenvectors in the columns transposed so that the eigenvectors are now in the rows, with the most significant eigenvector at the top, and Row Data Adjust is the mean-adjusted data transposed, i.e., the data items are in each column, with each row holding a separate dimension. [2]

*2.T-Distributed Stochastic Neighbor Embedding (t-SNE):*
t-SNE is a dimensionality reduction technique which tries to convert data point nearby into clusters and sends points that are beyond threshold to a very far distance. Let x correspond to the data point in the high dimensionality space and y denote the data points corresponding to the low dimensionality space. Then we find the conditional probability between the points denoted by $p_{i/j}$.

$$p_{j/i} = \frac{e^{\frac{-d_{ij}^2}{2\sigma_i^2}}}{\sum_k e^{\frac{-d_{ik}^2}{2\sigma_i^2}}} \quad (8)$$

$$p_{ij} = \frac{p_{i/j} + p_{j/i}}{2n} \quad (9)$$

Where $d_{ij}$ represents distance of $j^{th}$ feature from $i^{th}$ feature and $\sigma_i$ is the Gaussian variance centered at $i^{th}$ feature then $p_{ij}$ is found in the similar way conditional probability of low dimensionality features is represented as $q_{ij}$. we choose $q_{ij}$. In such a way that the resulted cost function is minimum

$$q_{j/i} = \frac{e^{-d_{ij}^2}}{\sum_k e^{-d_{ik}^2}} \quad (10)$$

$$cost = \sum_i KL(P_i||Q_i) = \sum_i \sum_j \log \frac{p_{j/i}}{q_{j/i}} \quad (11)$$

$$\frac{\partial Cost}{\partial \mathbf{y}_i} = 2\sum_j (\mathbf{y}_j - \mathbf{y}_i)(p_{j|i} - q_{j|i} + p_{i|j} - q_{i|j}) \quad (12)$$

*D.Classifier:*
A machine learning task that deals with identifying the class to which an instance belongs. we are going to compare four classifiers for our speaker identification.

*1. Feed Forward Neural Network:*
Feed Forward neural network we used consists of 2 hidden layers each hidden layer consists of large number of units each unit is connected to all the units in the next layer but none of them are inter connected each unit in the hidden layer is given a value called weight of the unit.

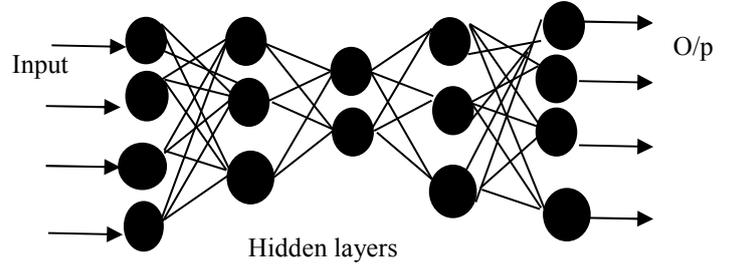

Figure 2.4  Feed forward network

*2. Support Vector Machine:*
This is a supervised machine learning algorithm. SVM can be used for classification and regression analysis. However, it is mostly used in classification problems. In this algorithm, we plot each data item as a point in n-dimensional space (where n is number of features you have) with the value of each feature being the value of a particular coordinate. Then, we perform classification by finding the hyper-plane that differentiate the two classes very well. Support Vectors are simply the co-ordinates of individual observation. Support Vector Machine is a frontier which best segregates the two classes (hyper-plane/ line).

*3. Decision Tree*
Decision trees, or Classification trees and regression trees, predict responses to the given data. To predict a response, we follow the decisions in the tree from the root (beginning) node down to a leaf node. The leaf node holds the response. Classification trees give responses that are insignificant, such as 'true' or 'false'. Regression trees give numeric responses.

*4. K- Nearest Neighbors*
This could be called straightforward extension of 1NN. we find the k nearest neighbour and do a majority voting. Classically k is odd when the number of classes is 2. A very popular thing to do is *weighted k-NN* where each point has a weight which is typically calculated using its distance.  This means that neighboring points have a higher vote than the farther points. It is quite obvious that the accuracy might increase when you increase k but the computation cost also increases.

## III. RESULTS AND DISCUSSION

The main goal of our project is to implement and compare different techniques in the conventional speaker recognition system and highlight the best algorithm that could be used to get efficient results. The main challenge in our project was the number of samples. Typically, according to the literature survey we did, we noticed that researchers had used samples as less as 3-4 to implement each algorithm. But in this project, we have used 15 speakers with each having three samples each and we have made an attempt to use 40-45 samples for checking each algorithm's performance which also tells us how the performance devalues as the number of sample increases. It gives us an insight on how one algorithm that proves to be the best for fewer samples turns out to be not very efficient with increase in speakers. We have tried to implement the above-mentioned algorithms in each block in every combination and permutation possible so that it gives a big picture about how the techniques could be exploited with varied number of speakers and requirement. These results tell us how the performance varies when the input speakers decrease in number while the best method chosen from the first set is used. So this paper on the whole extensively researches on every disparity that could lead to a different performance.

We are first comparing the performance of each combination along with the number of distinguishable speakers to find the best possible combination for speaker identification. The combination used in the first set is all three-feature extraction method with t-SNE and PCA. Table 3.1 shows the performance in percentage of each combination for 7 m speakers while using t-SNE for dimensionality reduction.

| t-SNE | MFCC (%) | LPCC (%) | PLP (%) |
|---|---|---|---|
| Complex Discrete Tree | 51.2 | 33.4 | 44 |
| Weighted Near Neighbor | 68.9 | 51.3 | 66.3 |
| Fine-SVM | 57.9 | 38 | 52.8 |
| Feed Forward | 51.5 | 47 | 50.3 |
| Bagged Trees-Ensemble | 67.4 | 47.3 | 66.2 |

Table 3.1 Performance of each frame

Above table shows performance of each frame of 7 speakers which are used for classification. Table 3.2 shows number of distinguishable speakers among the set of 7 speakers using this combination (t-SNE is used for dimensionality reduction).

| t-SNE | MFCC | LPCC | PLP |
|---|---|---|---|
| Complex Discrete Tree | 4 | 2 | 3 |
| Weighted Near Neighbor | 7 | 5 | 7 |
| Fine-SVM | 6 | 2 | 6 |
| Feed Forward | 4 | 3 | 5 |
| Bagged Trees-Ensemble | 7 | 5 | 7 |

Table 3.2 Distinguishable Speaker Among 7 Speakers

Table 3.3 and table 3.4 shows the performance in percentage of each combination for 7 speakers and number of distinguishable speakers among the chosen set of speakers respectively while using PCA for dimensionality reduction.

| PCA | MFCC(%) | LPCC(%) | PLP(%) |
|---|---|---|---|
| Complex Discrete Tree | 20.2 | 22.1 | 24 |
| Weighted Near Neighbor | 17.1 | 25 | 26.4 |
| Fine-SVM | 23 | 18.5 | 22 |
| Feed Forward | 18 | 17.5 | 20 |
| Bagged Trees-Ensemble | 13.7 | 18.6 | 22.4 |

Table3.3: Performance of each frame when PCA is used.

| PCA | MFCC | LPCC | PLP |
|---|---|---|---|
| Complex Discrete Tree | 2 | 1 | 2 |
| Weighted-Near Neighbor | 1 | 1 | 2 |
| Fine-SVM | 1 | 1 | 1 |
| Feed Forward | 2 | 1 | 1 |
| Bagged Trees-Ensemble | 1 | 1 | 1 |

Table3.4: Distinguishable speaker among 7 speakers (PCA)

From the above table, we clearly infer that t-SNE gives a better performance when compared to PCA as the number of speaker increases. As already mentioned the literature survey shows that PCA gives better results when the number of speakers involved was considerably less.

From the above tables, we can also interpret that t-SNE combination with weighted neighbor is the best possible combination for speaker verification. Figure3.1shown below shows how the performance of frames changes with the increase in number of speakers.

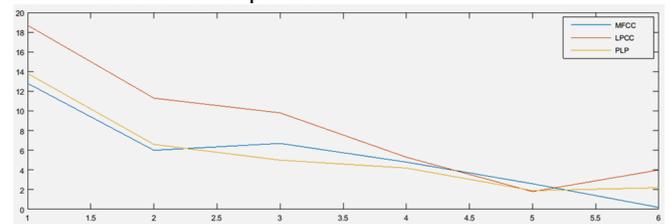

Figure 3.1 Performance vs Number of Speakers

From the figure 3.1 we can see that as the number of speakers increases performance of PLP feature extraction along with t-SNE dimensionality reduction followed by Weighted KNN gives a better performance than MFCC so if we are using a more number of speakers preferred identification combination is MFCC-tSNE-KNN but if we are planning for a speaker identification for a gadget with limited number of Speakers best preferred combination among our set is PLP-tSNE-KNN. Figure 3.2 shows the rate of change of performance as the number of speakers increase. More the stability of the graph, better is the performance of the combination, when the number of speakers is said to be changing.

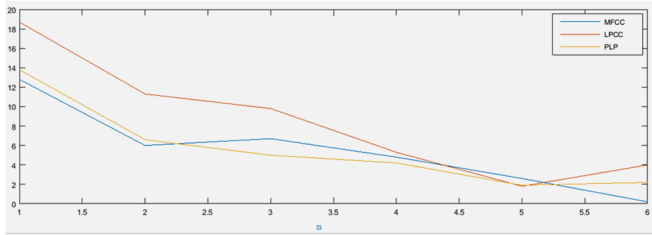
Fig3.2: Rate of change of performance versus speakers

From the above figure, we can see that rate of change in performance of PLP increases. But the rate of change of performance becomes minimum for MFCC, so for an unlimited number of users the performance becomes more stable in this case compared to PLP feature extraction method. From the results which are shown in the table it becomes very obvious that PLP-tSNE-KNN gives the best of performances when compared to the rest of the combinations. So, we implemented the mentioned combination for different set of speakers. The Table3.5 shows how the performance varies with varied number of users.

| Speaker >> | 2 | 3 | 4 | 5 | 6 | 7 |
|---|---|---|---|---|---|---|
| MFCC- KNN | 87.2 | 81.2 | 74.5 | 69.7 | 67.1 | 66.9 |
| LPCC- KNN | 81.3 | 70 | 60.2 | 54.9 | 53.1 | 49.1 |
| PLP– KNN | 86.2 | 79.6 | 74.6 | 70.4 | 68.5 | 66.3 |

Table 3.5: Performance of t-SNE, weighted K-NN combination for different number of speakers. The performance is represented in percentage.

Table3.5 shows the performance in percentage when different feature extraction algorithms are used in combination with t-SNE and weighted KNN for different number of speakers. it is observed that the reliability and efficiency in the performance increases with decrease in the number of speakers involved while training the network and it has already been pointed out before. Thus, reassuring the fact that training fewer number of speakers is easier. From the above table, we also infer that when it is fewer number of speakers for e.g. 2 speakers MFCC and PLP gives better performance.

Figure 3.6 shows Receiver Operating Characteristics of seven speakers using MFCC-t-SNE-KNN

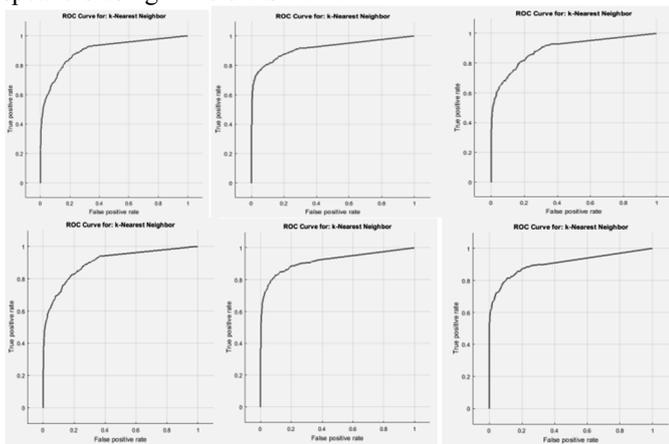
Figure 3.6 ROC of 2, 3, 4, 5 ,6 and 7 speakers of combination MFCC-tSNE-KNN

## IV. CONCLUSION

The paper is a comprehensive study of currently available algorithms for a speaker verification system. Our main observation is that the best combination varies depending on the input-number of speakers and the combination that performs best for fewer samples does not always give the best performance with larger number of samples . it is observed that the best combination for a large set, 30 samples of 7 speakers is MFCC-t-SNE-weighted KNN and for a  smaller set, 5-10 samples is MFCC/PLP-TSNE-weighted KNN. In conclusion, the best combination of algorithm must be chosen depending on the end requirement. Throughout the course of study a random dataset of male and female voices were used to train the network and an enhance study can be done by using either male voices or female voices or a combination of male and female voices in a definite proportion.